\documentstyle[aps,twocolumn,graphicx,floats,prl,amsmath,xspace]{revtex}

\newcommand{\bstar}{B$^\ast$\xspace}
\newcommand{\micron}{$\mu$m\xspace}

\begin{document}
\wideabs{
\title{Computer-Generated Holographic Optical Tweezer Arrays}

\author{Eric R. Dufresne$^{(1)}$, 
  Gabriel C. Spalding$^{(2)}$, 
  Matthew T. Dearing$^{(2)}$, 
  Steven A. Sheets$^{(2)}$, 
  and David G. Grier$^{(1)}$}

\address{$^{(1)}$Dept. of Physics, James Franck Institute, and
  Institute for Biophysical Dynamics \\
  The University of Chicago, Chicago, IL 60637\\
  $^{(2)}$Dept. of Physics, Illinois Wesleyan University, 
  Bloomington, IL 61702}

\date{\today}

\maketitle

\begin{abstract}
  Holographic techniques significantly extend the
  capabilities of laser tweezing, making possible
  extended trapping patterns for manipulating large
  numbers of particles and volumes of soft matter.
  We describe practical methods for creating arbitrary 
  configurations of optical tweezers using computer-generated
  diffractive optical elements.  While the discussion
  focuses on ways to create planar arrays of identical tweezers,
  the approach can be generalized to three-dimensional
  arrangements of heterogeneous tweezers and extended trapping
  patterns.
\end{abstract}

\pacs{ }

} 

\section{Introduction}

Since their invention in 1986 \cite{ashkin86}, optical
tweezers have become increasingly valuable tools for research in the 
biological 
\cite{biorefs}
and physical
\cite{physrefs}
sciences.
Using a focused beam of light to trap and move matter,
optical tweezers offer convenient, non-invasive access to processes
at the mesoscopic scale.
Most applications,  however,
have involved manipulating small numbers of particles
or small volumes of soft materials because
existing optical tweezer implementations can create
just a few tweezers at once.
Were they readily available,
large arrays of optical tweezers could be used
to organize microscopic particles into complex structures,
to sort them intelligently, to
study collective behavior in many-body systems,
and to manipulate materials too delicate to trap with a single tweezer.
We recently described \cite{dufresne98}
a method to create arrays
of optical tweezers using computer-generated
holographic beam splitters.  
This Article further explains how to design and fabricate
the necessary holograms and how to
integrate them into
\emph{holographic optical tweezer arrays} 
capable of trapping hundreds of particles simultaneously.

\begin{figure}[ht]
  \centering \includegraphics[width=3in]{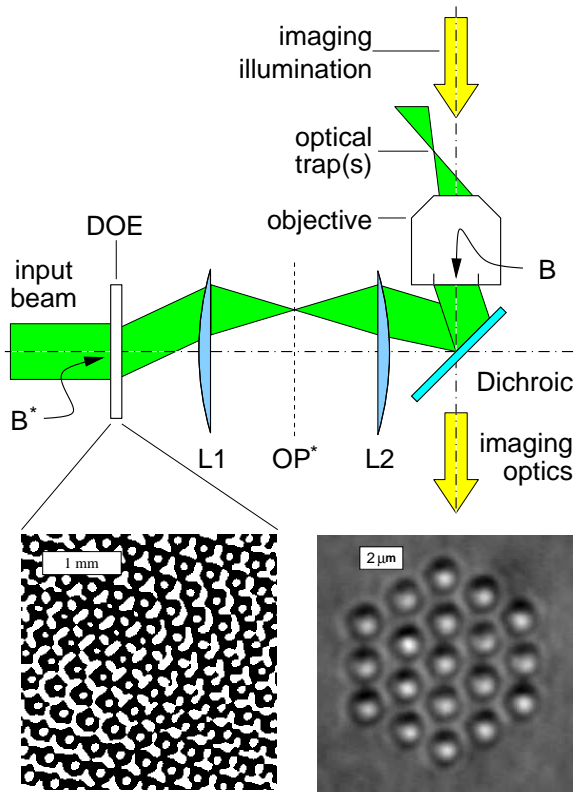}
  \vspace{1ex}
  \caption{Schematic representation of a typical holographic optical 
    tweezer array.  A collimated
    laser beam incident from the left is
    shaped by a diffractive optical element (DOE),
    transferred to an objective lens' back aperture (B)
    by lenses L1 and L2 and focused into a trapping array.
    OP$^\ast$ denotes the plane conjugate to the trapping plane.
    The point \bstar is conjugate to B.
    The phase pattern on the lower left (black regions shift
    the phase by $\pi$ radians)
    produced the traps shown in the lower right
    filled with 1~\micron diameter silica spheres
    suspended in water.}
  \label{fig:optics}
\end{figure}

\section{Optics of Optical Tweezers}

An optical tweezer traps particles with forces generated 
by optical intensity gradients.
Dielectric particles polarized by the light's electric field
are drawn up the gradients to the brightest point.
Reflecting, absorbing and low-dielectric particles, by contrast,
are driven by radiation pressure to the darkest point.
Optically generated forces strong enough to form a three-dimensional
trap can be obtained by
bringing a laser beam with an appropriately shaped
wavefront
to a tight focus with a high numerical aperture lens.
Microscope objective lenses offer an ideal combination of
minimal aberration and large numerical aperture and often
serve as the focusing element in practical implementations of
optical tweezers \cite{ashkin86} and variants
such as the optical vortex \cite{he95,vortex}.

The challenge in constructing an optical tweezer is to direct a
laser beam into the objective lens' back aperture
so that the beam fills the
aperture and so that its axis coincides with the optical
axis in the aperture's plane, 
at the point labelled B in Fig.~\ref{fig:optics}.
If the beam follows the optical axis, then it
comes to a focus and forms a trap in the center of the lens'
focal plane.  
If, on the other hand, it enters the back aperture
at an angle, the resulting trap is offset from the center of the focal
plane, as indicated schematically in Fig.~\ref{fig:optics}.

Directing the beam into the objective with a dichroic mirror allows
other wavelengths to pass through unimpeded and can be useful for
imaging the trapped particles, as in
Fig.~\ref{fig:optics}.  
The problem remains, however, of aiming
the beam.

The telescope formed by lenses L1 and L2 in Fig.~\ref{fig:optics}
addresses this problem by creating a conjugate point, 
\bstar, to the back aperture's center, B, at a
convenient location.
A beam of light passing through \bstar also passes through B and
forms an optical trap.  
In our implementation, L1 and L2 are high
quality plano-convex lenses with 250~mm focal lengths.  
Such long focal lengths help to minimize aberrations, particularly 
longitudinal spherical aberration, which would be detrimental to trapping
\cite{ke98,nemoto98,maianeto00}.
More compact optical trains would require additional attention
to minimizing wavefront distortions.
References~\cite{fallman97} and \cite{mio00} offer
more detailed discussions of this aspect of the optical design.

Multiple beams passing through \bstar all pass through B and thus all
form optical tweezers.  A diffractive optical element (DOE)
at \bstar, as shown in Fig.~\ref{fig:optics},
can split a single collimated
laser beam into any desired distribution of beams, each
emanating from \bstar at a different angle, and thus each forming a
separate trap \cite{dufresne98}.  
Figure~\ref{fig:optics}
shows the computer-generated pattern for a binary 
phase hologram together
with a photomicrograph of colloidal particles trapped in the resulting
array of optical tweezers.  
The remainder of this Article addresses the theory
and practice of creating holograms such as the example in
Fig.~\ref{fig:optics} suitable for projecting arbitrary arrangements
of optical tweezers.

\section{Holographic Tweezer Arrays}

\begin{figure}[b]
  \centering \includegraphics[width=3in]{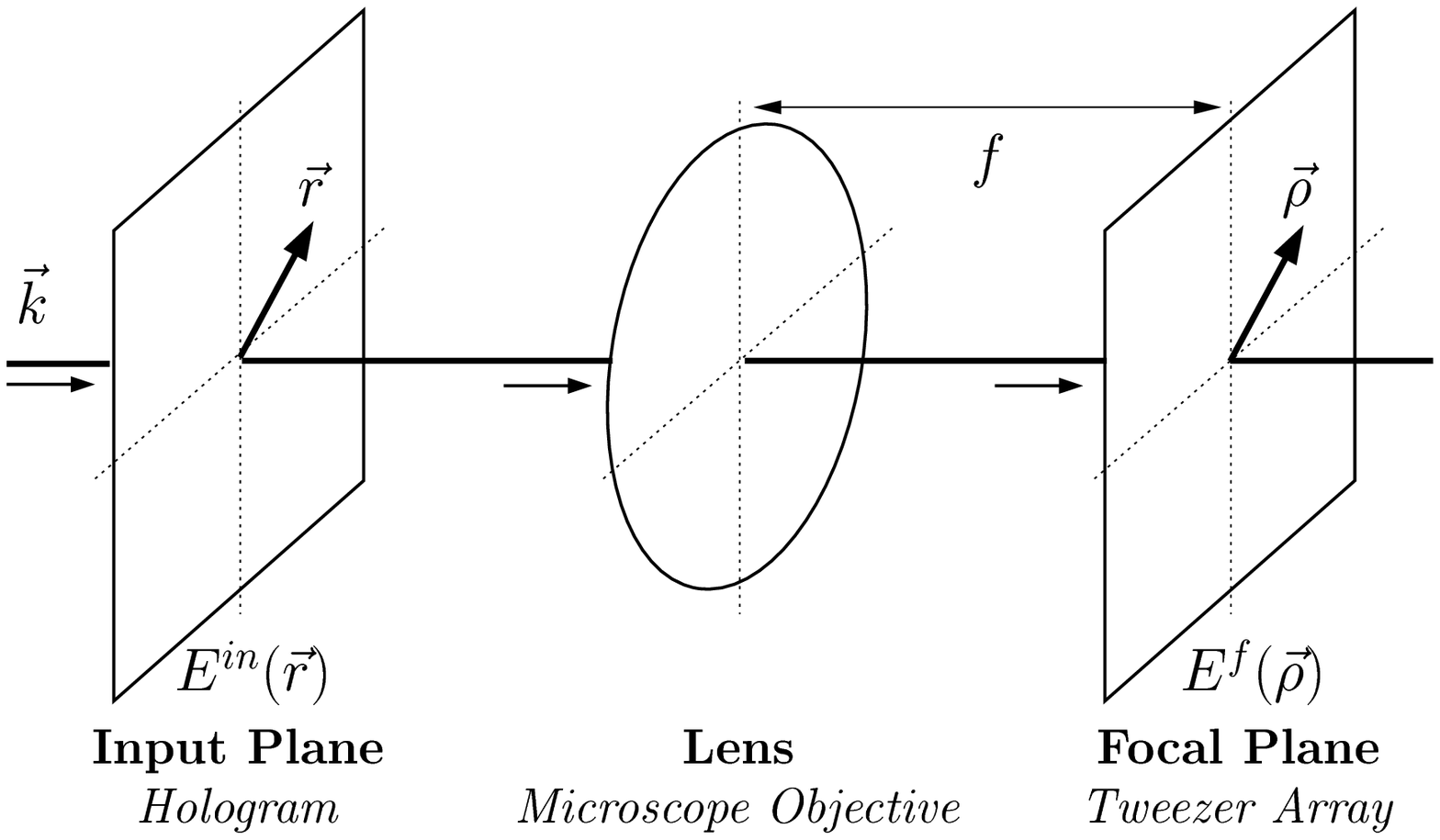}
  \vspace{1ex}
  \caption{Schematic representation of the 
    optical train highlighting the relationship between the beam geometry in 
    the input and focal planes. Monochromatic light, with wavevector 
    $\vec k$, is incident on the input plane.  A lens of focal length 
    $f$ projects the Fourier transform of the incident light's
    wavefront onto the focal plane. }
  \label{fig:fouropt}
\end{figure}
\subsection{Fourier Optics}
A planar array of optical tweezers can be described by the intensity
distribution, $I^f(\vec \rho)$, of laser light in the 
focal plane of a microscope's objective lens.
This pattern
is determined by the
electric field of light incident at its input plane, as
depicted in Fig.~\ref{fig:fouropt}.
Suppose that the input plane is illuminated by
monochromatic light of wavelength $\lambda$.
Its wavefront at the input plane, $E^{in}(\vec r)$, contains
both phase and amplitude information,
\begin{equation} 
  E^{in}(\vec r) = A^{in}(\vec r) \exp[i \Phi^{in}(\vec r)], 
\end{equation} 
where the amplitude, $A^{in}(\vec r)$, and phase, $\Phi^{in}(\vec r)$,
are real-valued functions.  
The electric field in the
focal plane has a similar form,
\begin{equation} 
  E^f(\vec \rho) = A^f(\vec \rho) \exp[i \Phi^f(\vec \rho)],
\end{equation} 
so that $I^f(\vec \rho) = |E^f(\vec \rho)|^2 = |A^f(\vec \rho)|^2$.
These fields
are related by the Fourier transform pair
\begin{align}
  E^f(\vec \rho) & = \frac{k}{2\pi f} \, e^{i \theta(\vec \rho)} \,
  \int d^2 r \, E^{in}(\vec r) \, e^{-ik \vec r \cdot \vec \rho/f} \\ 
  & \equiv {\cal F} \{ E^{in}(\vec r) \}, \qquad \mathrm{and}
  \label{eq:inf} \\
  E^{in}(\vec r) & = \frac{k}{2 \pi f} \, \int d^2 \rho \, 
  e^{-i\theta(\vec \rho)} E^f(\vec \rho) \, 
  e^{ik \vec r \cdot \vec \rho/f} \\ 
  & \equiv {\cal F}^{-1}\{E^f(\vec \rho) \},
  \label{eq:fin}
\end{align}
where $f$ is the focal length of the lens and $k = 2\pi/\lambda$ 
is the wavenumber
of the incident light.  
The additional
phase profile, $\theta(\vec \rho)$,
due to the lens' geometry 
does not contribute 
to $I^f(\vec \rho)$
and may be ignored without loss of
generality \cite{lipson81}.

\subsection{Phase-Only  Holograms}
Obtaining a desired wavefront in the focal plane requires
introducing the appropriate wavefront in the input plane.
Most lasers, however, provide only a fixed wavefront,
\begin{equation}
  \label{eq:input}
  E_0(\vec r) = A_0(\vec r) \, \exp [i \Phi_0(\vec r)].
\end{equation}
Shaping $E_0(\vec r)$ into $E^{in}(\vec r)$ involves
modifying both the amplitude and phase at the input plane.
Changing the amplitude with a passive optical element necessarily
diverts power from the beam and diminishes
trapping efficiency.
Fortunately, optical trapping relies on the beam's intensity
and not on its phase.
We can exploit this redundancy by setting
$A^{in}(\vec r) = A_0(\vec r)$ and modulating only the phase of
the input beam to obtain the desired trapping configuration.

Several techniques are available for achieving the necessary
phase modulation, and some of the associated 
practical considerations are
discussed in Section~\ref{sec:fabrication}.
For the purposes of the present discussion, we will refer
to the phase modulating element as a hologram or a diffractive
optical element and treat it as if it acts in transmission,
as shown in Fig.~\ref{fig:optics}.

After passing
through a phase modulating hologram, the electric field in
the input plane has a modified wavefront
\begin{equation}
  E^{in}(\vec r) = E_0(\vec r) \, \exp [i \Phi^{in}(\vec r) ],
\label{eq:holoin}
\end{equation}
where $\Phi^{in}(\vec r)$ is the imposed phase profile.
Calculating the phase hologram,
$\Phi^{in}(\vec r)$, needed to project a desired pattern of traps
is not particularly straightforward, as a simple
example demonstrates.

In a typical application of holographic optical tweezer arrays,
the undiffracted beam, $E_0(\vec r)$, 
projects a single optical tweezer into the
center of the focal plane with output wavefront 
$E^f_0(\vec \rho)$, and the goal
is to create displaced copies of this tweezer in the focal plane.
One possible wavefront describing an array of $N$ optical tweezers
at positions $\vec \rho_i$ in the focal plane is a superposition
of single (non-overlapping) tweezers
\begin{equation}
  E^f(\vec \rho) = \sum_{i = 1}^N \alpha_i \,
  E^f_0(\vec \rho - \vec \rho_i),
\end{equation}
where the normalization $\sum_{i=1}^N |\alpha_i|^2 = 1$ conserves energy.
$E^f(\vec \rho)$
may be written as a convolution 
\begin{align}
  E^f(\vec \rho) & = 
  \int d^2 \rho' \, E^f_0( \vec \rho' ) \, T(\vec \rho - \vec \rho') \\
  & \equiv E^f_0 \circ T(\vec \rho)
  \label{eq:convtwz2}
\end{align}
of $E^f_0(\vec \rho)$ with a
lattice function
\begin{equation} 
  T(\vec \rho) = \sum_{i=1}^N \alpha_i \, 
  \delta^{(2)}(\vec \rho - \vec \rho_i).
\end{equation}

Equations~(\ref{eq:fin}) and (\ref{eq:holoin})
relate $E^f(\vec \rho)$ to the associated
input wavefront: 
\begin{multline}
  E^{in}_0(\vec r) \, \exp [i \Phi^{in}(\vec r) ] 
  = {\cal F}^{-1} \left\{ E^f_0 \circ T(\vec \rho) \right\} \\
  = \frac{2\pi f}{k} \, {\cal F}^{-1} \left\{ E^f_0(\vec \rho) \right\} \,
  {\cal F}^{-1} \left\{ T(\vec \rho) \right\},
  \label{eq:dunno}
\end{multline}
by the Fourier convolution theorem.
The phase modulation needed to achieve the array of optical tweezers
then follows from Eq.~(\ref{eq:holoin}):
\begin{equation}
  \exp [i \Phi^{in}(\vec r) ] = 
  \frac{2\pi f}{k} \, {\cal F}^{-1} \left\{ T(\vec \rho) \right\},
\end{equation}
independent of the form of the single tweezer.

The phases of the complex weights, $\alpha_i$, must be selected so
that $\Phi^{in}(\vec r)$ is a real-valued function.
Unfortunately, the resulting system of
equations has no analytic solution.
Still greater difficulties are encountered in designing
more general systems of optical traps, including tweezers which
trap out of the focal plane or mixed arrays of conventional
and vortex tweezers.
Rather than deriving solutions for particular
tweezer configurations, we have developed more general
numerical methods
which we apply in 
the following Sections to creating planar arrays
optical tweezers.

\section{Adaptive-Additive Algorithm}
\begin{figure}[t]
  \centering \includegraphics[width=3in]{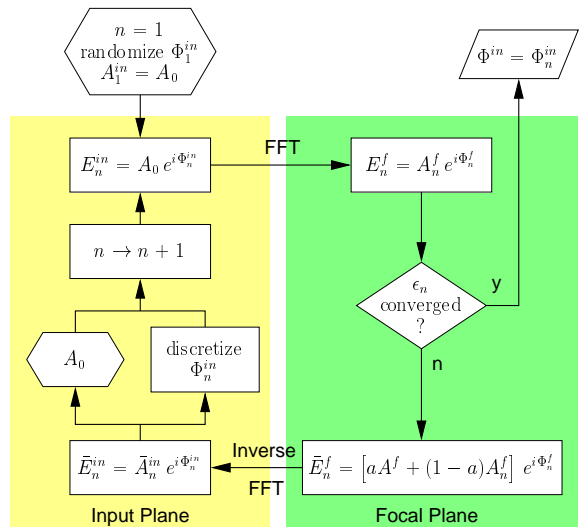}
  \vspace{1ex}
  \caption{Flow chart for the adaptive-additive algorithm.
    The phase modulation, $\Phi^{in}_n(\vec r)$, can
    be quantized into discrete steps with every iteration,
    as shown, or after the algorithm has converged.}
  \label{fig:algo}
\end{figure}
Our approach is based on the
adaptive-additive (AA) algorithm of Soifer \emph{et al.}
\cite{soifer97}, an iterative numerical technique which explores the
space of degenerate phase profiles, $\Phi^f(\vec \rho)$, to find a
phase modulation of the incident laser beam encoding any desired
intensity profile in the focal plane.
To facilitate calculation and fabrication, 
both the input and output planes are
discretized into $M \times M$ square arrays of pixels.
Optimal spatial resolution requires pixels in the
focal plane to be one half-wavelength on a side,
$\delta^f = \lambda/2$.
The number, $M$, of pixels on a side then depends on
the desired dimensions of the trapping array.
Lengths in the input and focal planes are
related by Eqs.~(\ref{eq:inf}) and (\ref{eq:fin}), so
that the corresponding pixel size
in the input plane is
$\delta^{in} = \lambda f / (M \delta^f) = 2f/M$.
If $\delta^{in}$ is inconveniently small, then L1 and L2
can be chosen so that a more 
amenable pixel size at \bstar corresponds to $\delta^{in}$
at B.

The AA algorithm, depicted in Fig.~\ref{fig:algo}, 
starts with an arbitrary initial guess
for $\Phi^{in}_1(\vec r)$ and an initial input wavefront
$E^{in}_1(\vec r) = E_0(\vec r) \, \exp[ i \Phi^{in}_1(\vec r)]$.
The Fourier transform of this wavefront is the starting
estimate for the output electric field:
$E^f_1(\vec \rho) = {\cal F}\{E^{in}_1(\vec r)\} = 
A^f_1(\vec \rho) \, \exp[i\Phi^f_1(\vec \rho)]$.
The corresponding intensity in the output plane,
$I^f_1(\vec \rho) = |A^f_1(\vec \rho)|^2$ is unlikely to be a 
good rendition of the desired intensity pattern, 
$I^f(\vec \rho) = |A^f(\vec \rho)|^2$.
The error,
\begin{equation} 
  \epsilon_1 \equiv
  \frac{1}{M^2} \sum_{i=1}^{M^2}
  [I^f(\vec \rho_i)-I^f_1(\vec \rho_i)]^2,
\end{equation}
is reduced by mixing a proportion, $a$, of the
desired amplitude into the field in the focal plane:
\begin{equation}
  \bar E^f_1(\vec \rho) = [ a A^f(\vec \rho) + (1-a)A^f_1(\vec \rho) ] \,
  \exp [i \Phi^f_1(\vec \rho)].
\end{equation}
Inverse transforming $\bar E^f_1(\vec \rho)$ yields the
corresponding field in the input plane,
$\bar E^{in}_1(\vec r) = \bar A^{in}_1(\vec r) \, 
\exp[i \Phi^{in}_2(\vec r)]$.
At this point, the amplitude in the input plane
no longer matches the actual laser profile, so
we replace  $\bar A^{in}_1(\vec r)$ with
$A_0(\vec r)$.
The result is an improved estimate for the 
input field:
$E^{in}_2(\vec r) = E_0(\vec r) \,\exp[i \Phi^{in}_2(\vec r)]$.
This completes one iteration of the AA algorithm.
Subsequent iterations lead to monotonically improving estimates,
$\Phi^{in}_n(\vec r)$, 
for the desired phase modulation
\cite{soifer97}.
The cycle is
repeated until the error, $\epsilon_n$, in the $n$-th iteration
converges to within an acceptable tolerance:
$(\epsilon_n - \epsilon_{n-1}) / \epsilon_n < \chi$.

The phase and amplitude fields are computed as arrays
of double-precision numbers, and their Fourier transforms calculated
with fast Fourier transform (FFT) routines.
Starting from  random input phases,
$\Phi^{in}_1(\vec r_i)$, uniformly distributed in the range 0 to $2 \pi$,
the AA algorithm typically requires eight iterations
to converge within $\chi = 10^{-6}$ of
an acceptably accurate local minimum of $\epsilon_n$
using an intermediate value for the mixing parameter, $a = 0.5$.

\section{Practical Considerations}

The AA algorithm generates phase profiles, $\Phi^{in}(\vec r)$,
that vary continuously between 0 and $2\pi$.
Actually creating a phase element with continuously varying phase
delay is difficult; usually only a small number of discrete levels
are available.  
Discretizing the output of the adaptive-additive
algorithm necessarily introduces errors.
These can be minimized by integrating the discretization step into the
AA algorithm itself, as shown in Fig.~\ref{fig:algo},
although this can lead to problems with convergence.

\subsection{Binarization}
\begin{figure}[h!]
    \centering \includegraphics{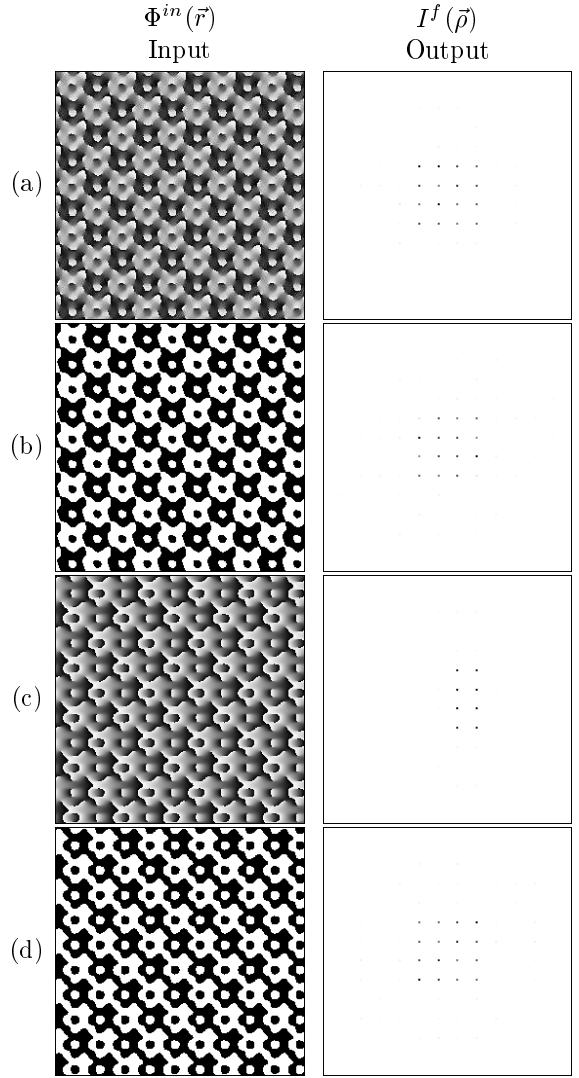}
  \vspace{1ex}
  \caption{Inversion symmetry in binarized holograms.  
    (a) A continuous hologram encoding a $4\times4$ array of tweezers.  
    (b) The binary version generates an array with
    missing tweezers. 
    (c) A continuous hologram encoding a $4\times 2$ array of tweezers. 
    (d) The binary version of hologram (c) makes a 
    satisfactory $4 \times 4$ array of tweezers.}
  \label{fig:zrohlf}
\end{figure}
   
The most straightforward phase
modulators
offer just two levels of phase delay, and are known as binary
holograms.
Beyond quantization errors and their attendant loss of efficiency,
binarization also imposes
inversion symmetry on the output wavefront, 
$E^f(\vec \rho) = E^f(-\vec \rho)$, and so limits what
patterns can be generated.
This might not seem a problem for inversion-symmetric patterns,
but interference between two sides of the pattern can
lead to unsatisfactory results, as shown in 
Figs.~\ref{fig:zrohlf}(a) and (b).
If, however, we anticipate the reflection and
calculate a phase mask encoding only half of the array, we
achieve much better results, as shown in Figs.~\ref{fig:zrohlf}(c)
and (d).
In practice, we repeat this calculation about twenty times
and choose the binary hologram with the best performance.

\subsection{Tiling}
\begin{figure}[htbp]
  \centering \includegraphics{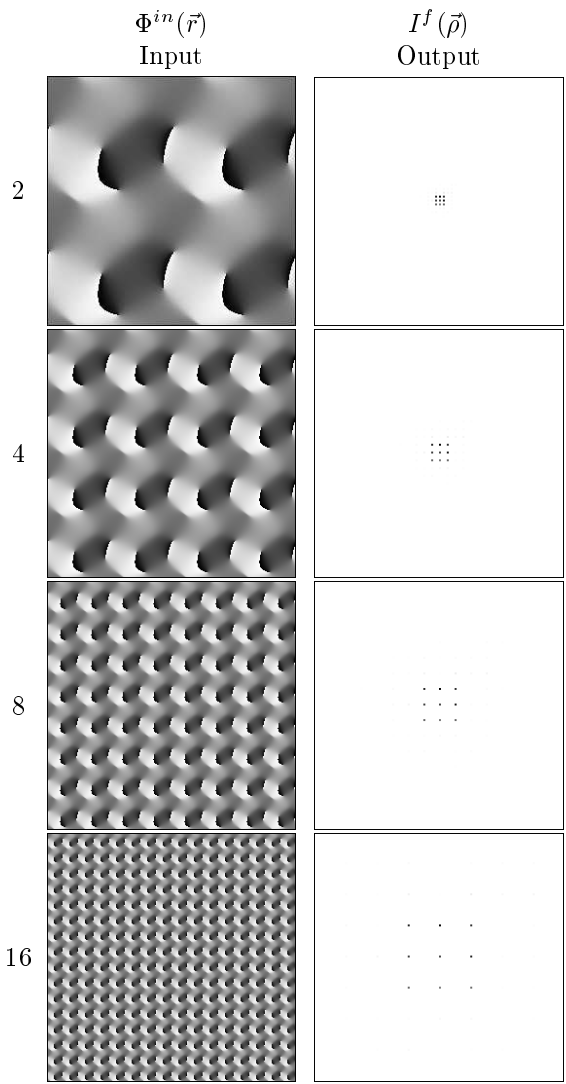}
  \vspace{1ex}
  \caption{Tiling a hologram encoding a $3 \times 3$ array
    of tweezers scales the spacing between tweezers without
    sacrificing resolution.  Marginal numbers indicate the number
    of copies tiled into each side.}
  \label{fig:scl}
\end{figure}

Because fast Fourier
transforms yield periodic functions, all holograms calculated
with the AA algorithm can be tiled smoothly.  
That is, they can
serve as the unit cell for new holograms
without introducing phase discontinuities at the unit
cell boundaries.  
The result of such tiling is to increase the spacing between
tweezers by an amount proportional to the number of tilings
along each dimension
without reducing the resolution or trapping ability
of the individual tweezers.
Fig.~\ref{fig:scl} shows successive tilings of a
hologram that generates
a $3 \times 3$
square array of tweezers, each labelled by
the number of unit cells tiled
along each side of the hologram.
The same number describes the relative spacings of the 
resulting tweezers.

We use this property to design holograms 
encoding tweezer arrays with large inter-tweezer spacings.
Increasing an array's lattice constant requires smaller features
in the input plane.
In order to resolve these small features, the
hologram's pixel size must be reduced.  Since the width of the
hologram is fixed, the number of pixels increases with the inverse
square of the pixel size.
Calculating such holograms can become computationally expensive.
Instead of directly calculating the hologram for
a desired lattice constant, therefore,
we calculate the smaller hologram encoding
the same pattern with a proportionately smaller
lattice constant.
This hologram can be
tiled to create a hologram for the desired tweezer spacing.
Tiling can be done either numerically or
physically, via a step and repeat mask fabrication process.

\section{Fabrication}
\label{sec:fabrication}

Phase profiles
can be recorded in the surface topography of an optical element 
\cite{swanson89},
or in controlled variations in a dielectric's index of refraction \cite{he95}.
Liquid crystal displays also have been used as
phase-modulating elements \cite{igasaki99}, and dynamically
reconfigurable tweezer arrays
have been demonstrated in principle
\cite{reicherter99}, although not yet in practice.
Some photorefractive elements such as those being explored as
optical memory devices also can be reconfigured,
but must be programmed optically.
Few, if any, are available as commercial optical elements.
Photorefractive holograms created with photographic
techniques \cite{he95} promise the greatest flexibility
for creating static tweezer arrays at very low cost, 
but do not appear to have advanced
beyond the research stage.

\begin{figure}[htbp]
  \centering \includegraphics[width=3in]{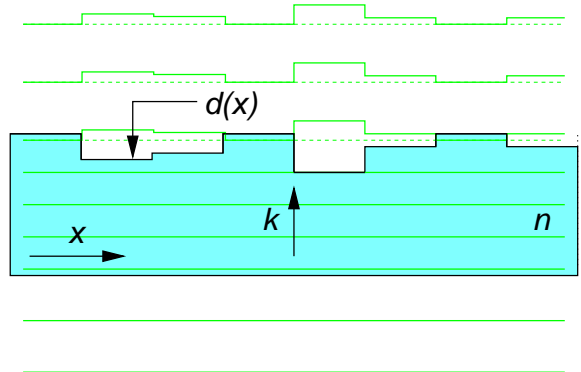}
  \vspace{1ex}
  \caption{Encoding phase in surface profile. 
    A plane wave incident upon the flat side of
    a transparent dielectric material acquires a spatially
    modulated phase upon passing through its textured
    surface.}
  \label{fig:phaseret}
\end{figure}

Surface patterning takes advantage
of well-established photolithographic techniques and can
be implemented easily and inexpensively.
We have taken this approach in creating our own
holographic optical tweezer arrays.
Fig.~\ref{fig:phaseret} shows the principle.
Light propagates more slowly in a dielectric material
than in air.
When a wavefront first enters the material, it is uniformly
slowed to a speed $c/n$, where $c$ is the speed of light in
vacuum and $n$ is the material's index of
refraction.
Parts of the wavefront emerging first from the textured surface
propagate at speed $c$, while sections
remaining in the material fall behind, picking
up a phase delay proportional to the extra thickness
of material.
Consequently,
the relative phase at $\vec r$ is proportional to the surface's
relief, $d(\vec r)$:
\begin{equation} 
  \Phi^{in}(\vec r) = 2 \pi (n-1) \frac{d(\vec r)}{\lambda}.
  \label{eq:phaseret} 
\end{equation}
A similar principle applies when imposing a pattern of phase delays
through the relief on a reflective surface, but with the
factor $n-1$ replaced by 2.

The pattern of hills and valleys needed to create a desired
phase profile can be formed in photoelastic polymer gels.
Such materials provide the recording medium
for commercial holographic printers.
These are not so common as photolithographic facilities
for surface etching, however, so we digress in the next Section
to describe the
details of our fabrication process.

\subsection{Reactive Ion Etching of Fused Silica}

\begin{figure}[htbp]
  \centering \includegraphics[width=3in]{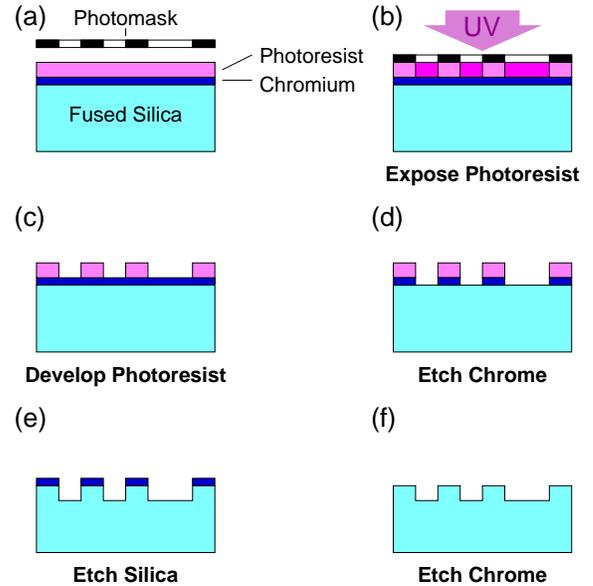}
  \vspace{1ex}
  \caption{Fabricating holograms with reactive ion etching.}
  \label{fig:fab}
\end{figure}

We employ reactive ion etching to create binary
holograms in 1-mm thick substrates of
polished fused silica, a transparent medium with
an index of refraction of $n = 1.456$ at the wavelength
of our trapping laser, $\lambda = 532$~nm.
Obtaining a phase shift of $\pi$ radians requires
a feature depth of
$\lambda/(2n-2) = 583$~nm.  
The fabrication process has
three main steps: creating a photomask, transferring this pattern
to an etch mask covering the silica, 
and then etching to a precise depth, Fig.~\ref{fig:fab}.

High contrast, high resolution film can be used
to create masks for many of the trapping patterns we have investigated.
We begin by laser printing the calculated phase profile
as a binary image, with black pixels representing a relative phase shift of 
$\pi$~radians, and white representing 0~radians.
This image is photoreduced 
to the actual dimensions of the hologram.
Each of our holograms covers a square whose width, $2f = 3.24$~mm,
matches the laser beam's
diameter at \bstar in Fig.~\ref{fig:optics}.
Holograms involving finer linewidths were vectorized
before processing with commercial mask writers at the
National Nanofabrication Facility.

We next create an etch mask on the surface of the
fused-silica substrate.
First the surface is protected with
a 25~nm layer of chromium and a 1.76~\micron
layer of positive photoresist, Fig.~\ref{fig:fab}(a).  
The photomask
is placed in contact with the photoresist, and the
entire sample is exposed to UV radiation, Fig.~\ref{fig:fab}(b).  
The photomask is removed and exposed regions of the photoresist are
dissolved away, revealing parts of the chromium layer,
Fig.~\ref{fig:fab}(c).  
Finally, the exposed chromium is removed with an acid wash, 
exposing the sections of silica to be etched, Fig.~\ref{fig:fab}(d).

Unprotected regions of the silica are susceptible to attack
by fluoride ions.
Reactive ion etching provides a controlled exposure to
ions generated by RF dissociation of a mixture of
oxygen and carbon tetrafluoride.
These reactive ions rapidly oxidize the organic photoresist, 
but are halted by the layer of metallic chromium.  
The unprotected regions of the silica surface
continue to be removed at a rate of about 0.5~nm/sec,
Fig.~\ref{fig:fab}(e), until the etched regions reach
the desired depth.
As the final step, we remove the remaining chromium to reveal
a precisely textured fused-silica surface,
Fig.~\ref{fig:fab}(f).

The etching process could be repeated with different photomasks
to produce a more nuanced pattern \cite{swanson89}.
$N$ such steps would yield $2^N$ gradations of phase delay.
Each step, however, would require planarizing and polishing
the previously etched pattern, recoating the surface, and
precisely aligning the new photomask over the existing pattern
before etching.
Not only is this is time consuming, it is
not necessary for many applications.
  
\subsection{Tolerances}
Regardless of the fabrication method, any practical
phase hologram will deviate from its design and these deviations
will degrade its performance.
We consider two principal fabrication defects: 
overall multiplicative error in the phase modulation due
to mismatches between wavelength and etch depth, and
random noise in the local phase shift due to roughness.
To quantify these defects' influence on hologram performance,
we define the efficiency, ${\cal E}$, to be the
fraction of incident laser power projected into
the planned tweezer pattern.  
For simplicity, we compare the intensity pattern in the
focal plane when the actual hologram is illuminated by a uniform
plane wave, $\tilde I^f(\vec \rho)$, to the ideal intensity pattern
in the focal plane, 
$I^f_0(\vec \rho) = (2 \pi f/k)^2 T^2(\vec \rho)$.  
The corresponding efficiency,
\begin{equation} 
  {\cal E} \equiv 
  \frac{\sum_{i=1}^{M^2} 
    T^2(\vec \rho_{i}) \tilde
    I^f(\vec \rho_{i}) }{\sum_{i=1}^{M^2} T^2(\vec \rho_{i})},
\label{eq:effdef}
\end{equation}
is a less stringent measure of the
agreement between the ideal and actual holograms than the error,
$\epsilon_n$, since it is possible to have ${\cal E} = 1$ when
$\epsilon_n > 0$, but $\epsilon_n = 0$ implies ${\cal E} = 1$.

To give a feel for the results obtained with our methods,
we calculate the efficiency of four standard holograms as a function
of the magnitude of the fabrication defects.
The four standard
holograms are continuous and binary versions of holograms
encoding $4 \times 4$ and $20 \times 20$ square tweezer arrays,
each with the same lattice constant.
We calculated all four holograms twenty times, and
selected the most efficient hologram from each group to use in the
the efficiency studies.

The phase modulation created by an etched hologram is
proportional to the etch depth, Eq.~(\ref{eq:phaseret}).  
If the etch rate is not precisely controlled, or if
the hologram is illuminated with
light of the wrong wavelength, the actual phase profile, 
$\tilde \Phi^{in}(\vec r)$, will differ from the design $\Phi^{in}(\vec r)$
by a scale factor, 
$\tilde \Phi^{in}(\vec r) = \alpha \Phi^{in}(\vec r)$.
\begin{figure}[htbp]
  \centering \includegraphics[width=3in]{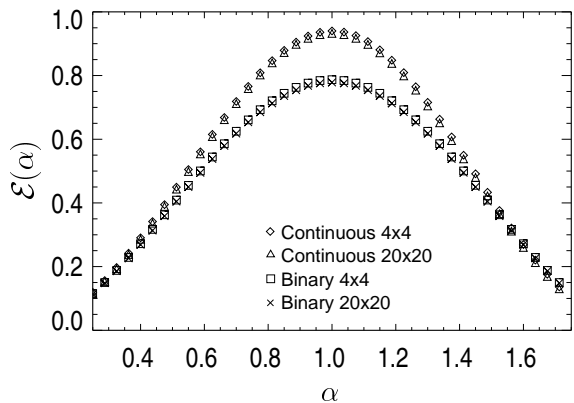}
  \vspace{1ex}
  \caption{Influence of phase errors on projection efficiency.
    Symbols indicate numerically calculated efficiencies for
    continuous and binary holograms encoding $4 \times 4$
    and $20 \times 20$ square arrays of tweezers.}
  \label{fig:edpth}
\end{figure}
As $\alpha$ departs from unity, most of the laser light not
contributing to the tweezer array is focused at the central
undiffracted spot.
Fig.~\ref{fig:edpth} shows the efficiency of
the four standard holograms as a function of $\alpha$.
Even the continuous holograms with $\alpha=1$ are not perfectly
efficient because the AA algorithm rarely identifies
a globally ideal phase modulation.
Binary holograms are still less efficient, with ideal 
efficiencies near 80\%.  
Reassuringly, Fig.~\ref{fig:edpth} suggests that a hologram's efficiency
does not depend strongly on precisely matching etch depth to the light's
wavelength.

Reactive ion-etching creates a rough surface, 
whose asperities
add random fluctuations to the phase profile.
We measured the surface topography
of our fused silica wafers after etching and found a 
Gaussian distribution of
etch depths, with a standard deviation of 60~nm or 
$\pi/10$~radians at 532~nm illumination.
This roughness is laterally uncorrelated
down to length scales of less than 280~nm.

\begin{figure}[htbp]
  \centering \includegraphics[width=3in]{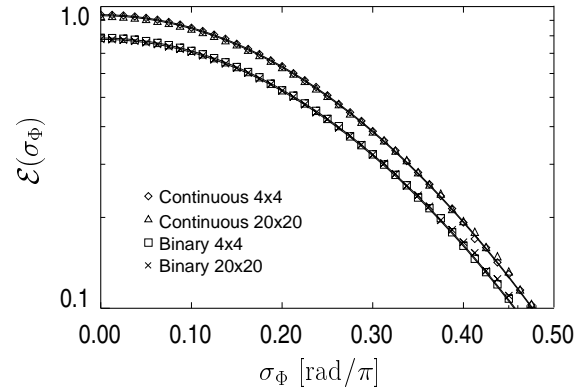}
  \vspace{1ex}
  \caption{Influence of roughness on efficiency.
    Symbols indicate numerically calculated efficiencies for
    continuous and binary holograms encoding $4 \times 4$ and
    $20 \times 20$ square arrays of tweezers subject
    to random Gaussian phase noise of magnitude $\sigma_\Phi$.
    Solid curves show the
    corresponding ensemble-averaged predictions from
    Eq.~(\protect\ref{eq:rough}).}
  \label{fig:rough}
\end{figure} 
We modeled such roughness by adding uncorrelated
Gaussian noise to the calculated phase profiles, 
\begin{equation} 
  \tilde \Phi^{in}(\vec r) = \Phi^{in}(\vec r) + \eta(\vec r),
  \label{eq:roughph}
\end{equation}
where the noise's probability distribution is given by
\begin{equation} 
  \rho(\eta) = \frac{1}{\sqrt{2 \pi \sigma_\Phi^2}}
  \exp \left(- \frac{\eta^2}{2 \sigma_\Phi^2} \right).
\label{eq:etadist}
\end{equation}
Fig.~\ref{fig:rough} shows how the efficiency of the four standard
holograms decreases with increasing surface roughness.

Combining
Eqs.~(\ref{eq:inf}) and (\ref{eq:roughph}) yields an
expression for the electric field profile in the focal plane
given a particular manifestation of 
the noise profile in the input plane,
\begin{equation} 
  \tilde E^f(\vec \rho) = \frac{k}{2 \pi f} \, \int d^2 r \,
  \exp \left( i \frac{k}{f} \, \vec r \cdot \vec \rho
    + i\Phi^{in}(\vec r) + i\eta(\vec r) \right).
\end{equation} 
Averaging
over all possible phase profiles 
yields
\begin{equation} 
  \langle \tilde E^f(\vec \rho) \rangle = 
  \exp \left( - \frac{1}{2} \, \sigma_{\Phi}^2 \right)
  E^f(\vec \rho),
\end{equation}
so that
\begin{equation}
  \langle {\cal E}(\sigma_{\Phi})\rangle = 
  {\cal E}(0) \exp \left(-\sigma^2_{\Phi}\right).
  \label{eq:rough}
\end{equation}
This result agrees well with numerically calculated efficiencies,
as can be seen in Fig.~\ref{fig:rough}.
Substituting the measured $\sigma^2_\Phi$ for our etched
binary holograms, we estimate that roughness diminishes their
efficiencies by a further 10\% to roughly 70\%.

\section{Further Considerations}
Using the techniques described above, we have created triangular
and square tweezer arrays which trap up to 400 particles at once.
Still larger arrays and less regular arrangements are certainly
feasible.
Even static holograms permit some degree of reconfigurability.
Rotating a hologram about its optical axis rotates the pattern
of tweezers in the plane.
Tilting it changes the aspect ratio.
Individual traps can be turned off by
blocking their beams in the plane conjugate to the
object plane, labelled OP$^\ast$ in Fig.~\ref{fig:optics}.
Such spatial filtering also can be useful for eliminating stray
laser light, and to block out any undiffracted portion of the
input beam.
Replacing lenses L1 and L2 with zoom lenses should permit
a degree of continuous scaling of the lattice constant.

The methods described in the previous Sections are appropriate
for projecting arrays of identical tweezers in the plane,
where each tweezer shares the properties of a single
tweezer formed by the unmodulated input beam.
Shaping the wavefronts of the individual beams, for example
to embed some optical vortices in an array of conventional
optical tweezers, requires a straightforward elaboration of
the AA algorithm \cite{soifer97}.
Creating three dimensional arrays, on the other hand, requires
more sophisticated calculations to avoid undesirable interference
effects, and will be discussed elsewhere.

\section{Acknowledgements}
This work was funded by the National Science Foundation
through grant DMR-9730189,
by the MRSEC program of the NSF through grant DMR-9888595 and by the 
David and Lucile Packard Foundation.


\end{document}